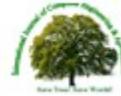

# MANAGING CONGESTION CONTROL IN MOBILE AD-HOC NETWORK USING MOBILE AGENTS

Ashish Kumar Mourya[1], Niraj Singhal[2]

[1]M.Tech. Scholar, Shobhit University, Meerut
[2]Associate Professor, Shobhit University, Meerut

**ABSTRACT:**

In mobile ad-hoc networks, congestion occurs with limited resources. The standard TCP congestion control mechanism is not able to handle the special properties of a shared wireless channel. TCP congestion control works very well on the Internet. But mobile ad-hoc networks exhibit some unique properties that greatly affect the design of appropriate protocols and protocol stacks in general, and of congestion control mechanism in particular. As it turned out, the vastly differing environment in a mobile ad-hoc network is highly problematic for standard TCP. Many approaches have been proposed to overcome these difficulties. Mobile agent based congestion control Technique is proposed to avoid congestion in ad-hoc network. When mobile agent travels through the network, it can select a less-loaded neighbor node as its next hop and update the routing table according to the node's congestion status. With the aid of mobile agents, the nodes can get the dynamic network topology in time. In this paper, a mobile agent based congestion control mechanism is presented.

## [I] INTRODUCTION

Mobile ad-hoc network (MANET) (or infrastructure less network) is a collection of mobile nodes which forms a network without central administration or standard support devices regularly available in conventional networks. Nodes in this network will both generate user and application traffic and carry out network control and routing protocols. In mobile ad hoc networks, congestion occurs with limited resources. Congestion occurs on shared networks when multiple users contend for access to the same resources (bandwidth, buffers, and queues). In such a network, packet transmission frequently suffered from collision, interference and fading, due to radio and dynamic topology. Congestion in mobile ad-hoc networks lead to transmission delays and packet loss, and causes wastage of time and energy for recovery. Congestion control refers to the network mechanism and techniques used to control congestion and keep the load below the networks capacity. Congestion handling can be divided into congestion recovery i.e. restore the operating state of the network when demand exceeds capacity and congestion avoidance i.e. anticipate congestion and avoid it so that congestion never occurs.

Routing protocols are network-layer protocols that are responsible for path determination and traffic switching. Routing can be classified into Adaptive Routing and Non-Adaptive Routing. In Adaptive Routing, routing decisions are taken for each packet separately i.e. for the packets belonging to the same destination, the router may select a new route for each packet. In it, routing decisions are based on condition or the topology of the network. In Non-adaptive Routing, routing decisions are not taken again and again i.e., once the router decides a route for the destination, it sends all packets for that destination on that same route. In it, routing decisions are not based on condition or the topology of the network.

Today, mobile agents are used in a number of research areas such as data communication systems (for example in activities related to monitoring, management, and security), distributed systems, computer languages, and intelligent systems. The deployment of mobile agents offers many advantages such as reduction of network traffic and latency, asynchronous execution, remote searching and filtering, efficient routing, inducing robustness, and fault tolerance.

Our focus is on the applications of mobile agents in data communication systems, in particular the field of resource management. The idea behind congestion control is allocating and managing resources in a network in order to attain best possible system performance. The ability of a mobile agent to exhibit autonomy, move freely across an arbitrary system, halt and resume its execution at any time offers some rewards in a highly distributed and heterogeneous environment. Migration through the nodes of a network and the potential to execute locally, which implies a sort of self-sufficiency with respect to global control, eliminates the need for strict centralized management and a lot of overhead.

The capacity of mobile agents to work in an asynchronous mode can provide efficient solutions for unreliable and low bandwidth network connections. Mobile agents have the potential to support mobile users in wireless network. In a volatile, prone to errors, limited connectivity and processing power settings, mobile users can disconnect while their agents roam through the network.

This paper presents a novel congestion control technique based on Mobile Agents.

## 2. RELATED WORK

This section presents a brief review of the work already done in this field. Kazuya Nishimura et al [1] have discussed a routing protocol that uses multi agents to detract network congestion for a Mobile Ad-hoc network. They have extended a dynamic routing protocol using mobile agent's protocol to be more generic, so that it can be effective in the combat of network congestion. Xiaoqin Chen et al [2] propose a congestion-aware routing metric which was employed data-rate, MAC overhead, and buffer queuing delay, with preference given to less congested high flow capacity links to improve channel utilization also they have proposed the Congestion Aware Routing protocol for Mobile ad-hoc networks (CARM). CARM has applied a link data-rate classification approach to prevent routes with mismatched link data-rates. CARM was only discussed and fictitious in relation IEEE 802.11b networks; however, it was applied to any multi-rate ad hoc network.

Consolee Mbarushimana et al [3], have exposed the performance of MANETs routing protocols is highly incumbent on the type of traffic generated or routed by mesne nodes. They have proposed a Type of Service Aware routing protocol (TSA), an enhancement to AODV, which uses both the ToS and conventional hop count as route selection metrics. TSA avoids congestion by distributing the load over a potentially greater area and therefore improving endemic reuse. Their simulation study reveals that TSA considerably improves the throughput and packet delay of both low and high priority traffic under different network operational conditions. Yung Yi et al [4] have evolved a fair hop-by-hop congestion control algorithm with the MAC constraint being imposed in the form of a channel access time constraint, using an optimization-based model. In the absence of delay, they have shown that their algorithm is globally stable using a Aleksandr Lyapunov-function-based approach. Next, in the presence of delay, they have shown that the hop-by-hop control algorithm has the property of endemic spreading. Also they have derived bounds on the "peak load" at a node, both with hop-by-hop control, as well as with end-to-end control, show that significant gains are to be had with the hop-by-hop scheme, and validate the analytical results with simulation.

Ming Yu et al [5] have proposed a link availability-based QoS-aware (LABQ) routing protocol for mobile ad hoc network based on mobility sortilege and link quality measurement, in addition to energy consumption estimate. They have provided highly reliable and better communication links with energy-competency. Yung Yi and Sanjay Shakkottai [6] have developed a fair hop-by-hop congestion control algorithm with the MAC constraint was being imposed in the form of channel access time compellable, using an optimization-based framework. In the absence of delay, they have shown that this algorithm was globally stable using a Lyapunov-function-based approach and in the presence of delay, they have shown that the hop-by-hop control algorithm has the property of spatial spreading.

Congestion in mobile ad-hoc network leads to transmission delay and packet loss, and causes wastage of time and energy on recovery. Routing protocols which are adaptive to the congestion status of a mobile ad hoc network can greatly improve the network execution. Xiaoqin chen et al [7] have proposed a congestion-aware routing protocol for mobile ad-hoc which has used a metric incorporating data-rate, MAC overhead, and buffer delay to encounter congestion. This metric was used, together with the avoidance of mismatched link data-rate routes, to make mobile ad-hoc networks durable and adaptive to congestion. RamaChandran and Shanmugavel [8] have proposed and studied three cross-layer designs among physical, medium access control and routing (network) layers, using Received Signal Strength (RSS) as cross-layer interaction parameter for energy patronage, unidirectional link rejection and reliable route creation in mobile ad-hoc network.

Sung-Ju Lee et al [9] has proposed an on-demand routing design called Split Multipath Routing (SMR). This establishes and utilizes multiple routes of maximally disjoint paths. The route recovery process is minimized and the message overhead is controlled by using multiple routes. The data packets are distributed into multiple paths of active sessions by using a per packet allocation scheme. Ducksoo shin al [10] has proposed A2OMDV which is an extension to AOMDV. The static route switching of AOMDV is resolved in A2OMDV. The best route among the multiple paths can be selected by the source node by maintaining the status of the node. In terms of throughput and delay, the A2OMDV shows a better performance during heavy loads.

N. Jaisankar et al [11] have proposed an AODV protocol for multipath routing. For the improvement of scalability, multiple routes were developed in this paper. The routing overhead incurred in maintaining the connection between source and destination nodes can be reduced by determining the multiple paths in a single route discovery. Due to the node mobility or battery failure, the primary path may fail. The secondary paths are used to transmit data packets so that extra overhead is generated by a fresh route discovery. Bo Xue et al [12] have proposed a Link Optimization Ad-hoc on-demand Multipath Distance Vector Routing (LOAOMDV). A new route reply process can be established using four bits information of the RREP packet. The omitted reverse path is established and the number of common node in several paths is reduced. The end-to-end delay is reduced, network lifetime is extended, packet timely can be reclaimed, and the ratio of packet chaotic sequence is reduced in this protocol.

Juan J. G´alvez et al [13] have proposed a spatially disjoint multipath routing protocol without location information. This protocol discovers multiple paths between two endpoints in one route discovery and it also measures the distance between the paths. The paths discovered by this route discovery mechanism traverse all geographical regions. Ash Mohammad Abbas et al [14] have discussed that the path diminution is unavoidable when a protocol discovers multiple node-disjoint paths in a single route discovery. Path diminution is mitigated using these schemes. Here they have proved that it is not possible to develop an efficient algorithm which guarantees computation of all node disjoint paths between a given pair of nodes in a single route discovery.

Kun-Ming Yu et al [15] have proposed a new protocol to improve existing on-demand routing protocols. When the network topology changes, the routing protocols construct multiple routing protocols and it can transmit data packets dynamically through backup routes. S.Karunakaran et al [16] have presented a Cluster Based Congestion Control (CBCC) protocol that consists of scalable and distributed cluster-based mechanisms for supporting congestion control in mobile ad hoc networks. The distinctive feature of their approach is that it is based on the self organization of the network into clusters. The clusters autonomously and proactively monitor congestion within its localized scope.

The next section provides proposed congestion control mechanism using mobile agents.

## 3. THE CONGESTION PROBLEM

In a network with shared resources, where multiple senders compete for link bandwidth, it is necessary to adjust the data rate used by each sender in order not to overload the network. Packets that arrive at a router and cannot be forwarded are dropped, consequently an excessive amount of packets arriving at a network bottleneck leads to many packet drops. These dropped packets might already have travelled a long way in the network and thus consumed significant resources. Additionally, the lost packets often trigger retransmissions, which mean that even more packets are sent into the network. Thus network congestion can severely deteriorate network throughput. If no appropriate congestion control is performed this can lead to a congestion collapse of the network.

Congestion is a major cause for packet loss in MANETs and reducing packet loss involves congestion control running on top of a mobility and failure adaptive routing protocol at the network layer. Congestion non-adaptive routing in MANETs may lead to the following problems:

◻◻Long delay: It takes time for a congestion to be detected by the congestion control procedure. In furious congestion situations, it may be better to use a new route. The problem with an on-demand routing protocol is the delay it takes to search for the new route.

◻◻High overhead: In case a new route is expected, it takes processing and communication effort to discover it. If multipath routing is used, though an alternate route is readily found, it takes effort to maintain multiple paths.

◻◻Many data packet losses: Many packets may have already been lost by the time congestion is detected. A typical congestion control solution will try to reduce the traffic load, either by decreasing the sending rate at the sender or dropping packets at the intermediate nodes or doing both. The consequence is a high packet loss rate or a small throughput at the receiver.

This paper presents how we control the congestion in ad-hoc network where path is already selected by the source node.

In a network with shared resources, where multiple senders compete for link bandwidth, it is necessary to adjust the data rate used by each sender in order not to overload the network. In multi-rate ad-hoc networks, throughput via a given route is limited by the minimum data-rate over its entire constituent links. Consider a route with figurer 1 significantly different data rates over each of its links (e.g. A-> B -> D -> F -> H). Let us call us such a route a mismatched data-rated route. When large-scale traffics (such as multimedia streams) are transmitted in such a route, the benefits of having multi rate links can be compromised. There is potential for congestion at any node which heads a link with a slower data rate than previous links, in a mismatched data rate route (e.g. node F in the example path), due to earlier high data rate node forwarding more traffic into low data rate nodes, long queue of data-packets may occurred on such paths. Clearly, avoiding, or at least lessening the mismatch in, multi data rate routes are important in combatting congestion.

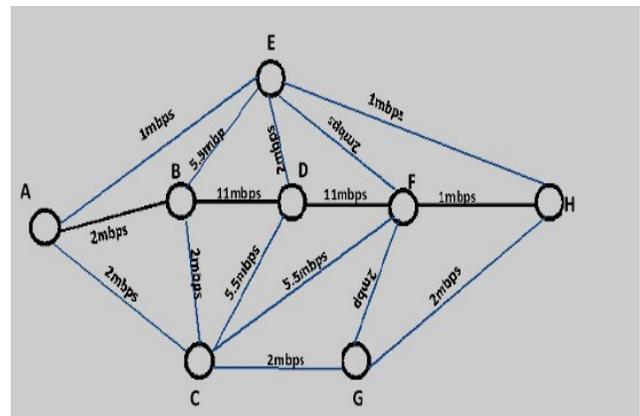

Figure 1: An example of 802.11b multi-rate Ad-hoc network

## 4. PROPOSED WORK

International Journal of Computer Engineering & Applications, Vol. IV, Issue I/III

www.ijcea.com ISSN 2321-3469 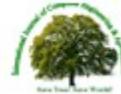

Congestion in wireless networks is slightly different from that of wired networks. The following are the two general cause of congestion:-

(i). The throughput of all nodes in a particular area gets reduced because many nodes within range of one another attempt to transmit simultaneously, resulting in losses.

(ii). The queue or buffer used to hold packets to be transmitted, overflows within a particular node. This is also the cause of losses.

Congestion adaptive routing has been investigated in several studies as presented in previous section. The approaches in all the cited studies converge in evaluating or assessing the level of activity in intermediate nodes by measuring either the load or the delay. Based on the gathered information, the optimal path is established trying to avoid the already or likely to become congested nodes. However, none of the research reported has evaluated the effect service type of the traffic carried by intermediate nodes has on the performance of routing protocols.

The route discovery process of most of MANETs routing protocols do not consider the status of their queues, before advertise themselves as candidate to route traffic to the destination. This might result into long delays or packet drops for newly arriving traffic, failing to be transmitted ahead of the already queuing traffic. The performance of the mobile ad hoc networks is strongly influenced by the congestion problem. A congestion control scheme consists of a routing algorithm and a flow control scheme. In earlier research, the routing and the flow control problems have been considered separately. To achieve better performance and better congestion control, the routing and the flow control must be considered jointly [17]

In this paper, we propose to design and develop agent based congestion control architecture. In this architecture, all the nodes are mobile and information about network congestion is collected and distributed by mobile agents (MA). Each node has a routing table that stores route information for every destination. MA starts from every node and moves to an adjacent node at every time. The MA updates the routing table of the node it is visiting.

The agent based congestion routing can be explained from the following figure:

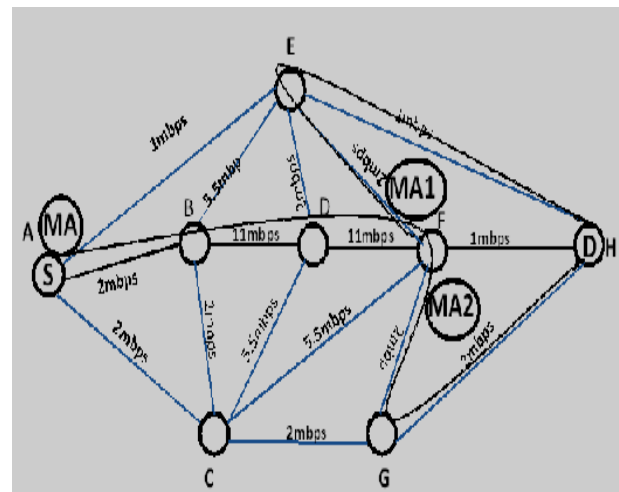

Figure 2: Mobile Agent based congestion control

In multi-rate ad-hoc networks, different data-rates will almost certainly lead to some routes having different links with quite different data-rates. If lower data-rate links follow higher data-rate links, packets will build up at the node heading the lower data-rate link, leading to long queuing delays

In this proposal, the node is classified in different categories depending on whether the traffic belongs to background, best effort, and video or voice data respectively. Then MA at each node estimates the congestion level for each traffic class by checking the queue status and a priority is assigned for the node based on the measured congestion level. Using this classification, a node with no traffic or with delay-insensitive traffic is considered more priority so that it can receive more traffic than a low priority node. The congestion level of every node is updated every time there is change in traffic type, and it is periodically propagated to neighbors.

To find the congestion control mechanism, we have followed these steps:-

Step 1: The sources 'S' check the number of available one hop neighbor, for example A-E, A-B, A-C).

Step 2: the Mobile agent selects the path to move towards the destination 'D' from source node S (as given in the Figure 2).

Step3: Now the mobile agent detect that congestion occurs between 'F-H' nodes due to earlier high data rates node (S-B-D-F) forwarding more traffic into low data rate node F.

Step 4: Now the source check the number of available one hop neighbors (E or G) and clones the Mobile Agent (MA) to that neighbor, MA1 & MA2.

Step 5: The MA1 moves towards the destination 'D' and node 'H' in a hop-by-hop manner in the path P1 and MA2 in P2 respectively. Then the MA1 calculate the data rates of that path P1 and similarly MA2 calculates the data rate of P2.

**P1 data rate = Data size / Channel delay (E node)**

**P2 data rate = Data size / Channel delay (G node)**

Step 6: Now the source selects path using highest date rates of P2(S-B-D-F-G-H) and send the data through the corresponding path.

## 5. CONCLUSION

This paper presents an agent based congestion control technique. In this technique, the information about network congestion is collected and distributed by mobile agents. A mobile agent starts from every node and moves to an adjacent node at every time. A node visited next is selected at the equivalent probability. The mobile agent brings its own history of movement and updates the routing table of the node it is visiting. The mobile agent calculates the data rate of their corresponding nodes and then selects the nodes that have highest data rate.